\documentclass[]{interact}

\usepackage{epstopdf}
\usepackage[caption=false]{subfig}

\usepackage[natbibapa,nodoi]{apacite}
\theoremstyle{plain}

\theoremstyle{definition}

\theoremstyle{remark}

\usepackage{siunitx}
\usepackage{pgf}
\usepackage{pgfmath}
\usepackage{xparse}
\usepackage{siunitx}
\newcommand{\signif}[1]{%
    \pgfmathparse{#1}%
    \ifdim \pgfmathresult pt < 0.0001pt %
        $\boldsymbol{\pgfmathprintnumber[assume math mode=true]{\pgfmathresult}}$***
    \else
        \ifdim \pgfmathresult pt < 0.001pt %
            $\boldsymbol{\pgfmathprintnumber[assume math mode=true]{\pgfmathresult}}$**
        \else
            \ifdim \pgfmathresult pt < 0.005pt %
                $\boldsymbol{\pgfmathprintnumber[assume math mode=true]{\pgfmathresult}}$*
            \else
                $\pgfmathprintnumber[assume math mode=true]{\pgfmathresult}$
            \fi
        \fi
    \fi
}

\usepackage{subcaption}
\usepackage{xcolor,colortbl}
\usepackage[edges]{forest}
\usepackage{amsmath}
\usepackage{siunitx}
\usepackage{graphicx}
\usepackage{lscape}
\usepackage{adjustbox}
\usepackage{pdflscape}

\graphicspath{{figures/}}

\newcommand{\pl}{PrairieLearn{}~\citep{west2015prairielearn}}
\begin{document}


\title{Mining Hierarchies with Conviction: Constructing the CS1 Skill Hierarchy with Pairwise Comparisons over Skill Distributions}

\author{
\name{Dip Kiran Pradhan Newar\textsuperscript{a}\thanks{CONTACT D. K. P. Newar. Email: dip.pradhannewar@usu.edu}, Max Fowler \textsuperscript{b}, David H. Smith IV\textsuperscript{b} and Seth Poulsen\textsuperscript{a}}
\affil{\textsuperscript{a} Utah State University, Logan, Utah, USA \textsuperscript{b}University of Illinois at Urbana-Champaign, Urbana, IL, USA}
}


\maketitle

\begin{abstract}
\textbf{Background and Context:} Some skills taught in introductory programming courses are categorized into 1) \textit{explaining} code, 2) arranging lines of code in correct \textit{sequence}, 3) \textit{tracing} through the execution of a program, and 4) \textit{writing} code from scratch. \newline
\textbf{Objective:} Knowing if a programming skill is a prerequisite to another would benefit teachers in properly planning the course and structuring the order in which they present activities relating to new content. Prior attempts to establish a skill hierarchy have suffered from methodological issues. \newline
\textbf{Method:} In this study, we used the conviction measure from association rule mining to perform pair-wise comparisons of five skills: Write, Trace, Reverse trace, Sequence, and Explain code. We used the data from four exams with more than 600 participants where students solved programming assignments of different skills for several programming topics. \newline
\textbf{Findings:} Our findings matched the previous finding that tracing is a prerequisite for students to learn to write code. Contradicting the previous claims, our analysis showed that using the mean threshold writing code is a prerequisite to explaining code. However, there is no clear relationship when we change the threshold to the median. Unlike prior work, we did not find a clear prerequisite relationship between sequencing code and writing or explaining code. \newline 
\textbf{Implications:} Our research can help instructors by systematically arranging the skills students exercise when encountering a new topic.
The goal is to help instructors properly teach and assess programming in a fashion most effective for learning by leveraging the relationship between skills. 
\end{abstract}

\section{Introduction}


Programming is the primary skill anyone pursuing a career in the Computer Science is expected to master. 
Every CS major student is taught the basics of programming in their introductory programming course (CS1) and expected to gain sufficient mastery in them to be successful in subsequent courses.

A seminal ITiCSE working group by \citet{mccracken2001report} found that, even after taking an introductory computing course, many students still struggle in constructing basic programs, a finding that has been replicated in a variety of contexts since~\citep{lister2004multi, whalley2009bracelet}.
These findings motivate the need to improve teaching methodologies within CS1 to foster stronger foundations in programming fundamentals.
Programming is often discussed as being composed of a set of ``skills'' that are defined by how students interact with existing code and write code themselves~\citep{xie2019theory, lister2006not}.
These can include: tracing, reverse tracing~\citep{hassan2021exploring}, sequencing~\citep{parsons2006parson}, explaining code~\citep{fowler2021should}, and writing code. 
Previously, researchers sought to find the relationship between skills and helping students learn programming skills using statistical analysis to see how tracing and explaining are related to writing code~\citep{lopez2008relationships, fowler2022reevaluating}. 
However, these studies sought to establish relationships between introductory programming skills using statistical methods that could only show correlations between skills, and therefore not truly illuminate prerequisite relationships.

In this paper, we use association rule mining to find the relation via pairwise comparisons of skills (in total, ten pairs). We seek to investigate the following research question:
\begin{itemize}
\item[\textbf{RQ:}] What prerequisite relationships exist among the skills of explaining, writing, tracing, reverse tracing, and sequencing code?
\end{itemize}
Using the conviction measure from association rule mining, we analyzed the data from a large public university in the United States from four exams given over the course of a 16 week semester to calculate the prerequisite relationship. Finally, we propose a skill hierarchy using the dependency of skills that shows relationships between these five skills. 


\section{Related Work}

Much of the work investigating programming skills descends from ``The McCracken Group'' and their analysis of students' proficiency with programming after CS1~\citep{mccracken2001report}. Their finding, that many students were far from reaching the expected standards of their programming courses, jump-started a long line of work aimed at determining the sequence in which students may learn programming skills. It may be the most meaningful way to inform the way we teach students how to program. To evaluate the findings from ~\citet{mccracken2001report},  ~\citet{lister2004multi} conducted studies with students with different nationalities to assess why students performed poorly in the McCracken study. They found that student lack of programming skills resulted in poor performance in the study. The development of a hierarchy structure of programming skills helps to refine the theories of programming skills by providing necessary guidance on ordering and prerequisites of the skills. For example, \citet{xie2019theory} proposed an often-cited theory of instruction for the skills used in introductory programming. Hierarchies are beneficial in supporting the development and refinement of such theories, especially if they can identify prerequisite relationships.

\subsection{The previous efforts at skill hierarchy determination}

There have been myriad efforts to understand the existence and strength of
possible programming skill hierarchies. Among them,
\citet{lopez2008relationships} are notable for presenting the first such
hierarchy. While their hierarchy contains multiple smaller skills, the
largest finding of their work was that tracing may be a prerequisite for reading
code, which may then be a prerequisite for writing code. Although this hierarchy
was promising, their sample size was relatively limited, leaving the hierarchy
without strong verification.

Multiple authors continued along the lines of this hierarchy.
\citet{lister2009further} successfully replicated the original
\citet{lopez2008relationships} hierarchy in a Python course and found that programming skills have no relation with programming language. Following, \citet{venables2009closer} directly replicated the original study with similar results but found that explaining and writing skills are sensitive to the type of question. ~\citet{sheard2008going} found ability to read code is correlated with writing skills, while ~\citet{kumar2015solving} found a strong correlation of tracing skills with writing code.

~\citet{yamamoto2011relationship} replicated the work of \citet{lopez2008relationships} using the Structural equation model and found that students ability to edit existing code and write new code from scratch depends on their ability to explain code. ~\citet{mendoza2018intervention} study shows the students code understanding skills have an impactful effect on students ability to write code. ~\citet{harrington2018tracing} in their study with CS2 students found that the gap between tracing and writing skills seems to disappear when a student masters the skills. ~\citet{corneyexplain} study with the CS2 students found a strong correlation between the students ability to code writing with explaining. \citet{pelchen2020evidence} found results similar to \citet{corneyexplain} and found that the results by~\citet{harrington2018tracing} could be affected by the difference in the difficulty of the tracing, writing and explaining questions.

Further work either investigating hierarchies or the links between skills has generally found similar relationships between skills, suggesting some continuity to the basic ideas behind a programming skill hierarchy. Notably, the precise strength of these relationships between skills differs by context and students' prior experience
~\citep{teague2012swapping,kikuchi2016investigating, yamamoto2012skill}.
 
However, most of the previous work in skill hierarchies remains largely
\textit{correlational} in nature. \citet{fowler2022reevaluating} raised this issue in their work, finding that while they \textit{could} replicate the original hierarchy from \citet{lopez2008relationships}, it was neither the only possible hierarchy nor necessarily the best hierarchy that appears when using purely correlational methods. 
Their results suggest both that these
skills are naturally tightly coupled, and thus hierarchies may be fluid, but
also that the methods used so far are not the best for determining the true
prerequisiteness of different pairs of programming skills.

Recent work by \citet{pelchen2020evidence} represents the first work in this
line that seeks to establish a hierarchy using measures that are not purely
correlational in order to properly handle the directionality of the relationships
between skills. They found more evidence for tracing serving as a prerequisite for explaining code, but did not identify other relationships, did not investigate code writing questions, and have a small data set with few questions per skill. Further, the work was carried out in a context assuming writing code as a terminal skill. In
contrast, we make no a priori assumptions about the relationships between
skills, and have a data set with enough questions per skill to allow us to make
generalizable claims about questions that target certain skills, rather than
just claims about individual questions.

We contribute to the search for strong evidence of prerequisiteness by applying
association rule mining techniques to a large set of student data, similar to
the data available to \citet{yamamoto2011relationship} 
and \citet{fowler2022reevaluating}.

\section{Methodology}

In this paper, we study five programming skills: write, trace, reverse trace, sequence, and explain and the relationships between them. 

\subsection{Data Collection}


\begin{table}
  \centering
  \begin{tabular}{ccl}
    \toprule
    Exam & Total Students & Types\\
    \midrule
    Exam1 & 677 & arithmetics, conditionals, strings, lists\\
    Exam2 & 649 & conditionals, loops\\
    Exam3 & 625 & patterns, advanced strings\\
    Exam4 & 611 & loops, patterns, objects, files, dictionaries\\
  \bottomrule
    \end{tabular}
    \caption{Dataset Summary}
  \label{tab: dataset}
\end{table}

We collected data from four exams from a non-major introductory Python class at a large public US institution. The class had a total of 677 students, although each exam was completed by a different number of students due to drops later in the semester. Table~\ref{tab: dataset} shows the summary of the dataset concerning the number of students who completed each exam and which topics each exam covered. 

All five of the skills measured had different question types. Code writing ability was assessed with small Python programming exercises, no more than a function in size. Tracing ability was measured by having students read code and determine the code's output. Each students are given a code snippet and are asked to answer the output generated from that code. As the name implies, reverse trace instead gave students an expected output and asked students to provide an input to the missing sequence in the code snippets that produces the shown output, in the style of ~\citet{hassan2021exploring} work. Code sequencing, properly sequencing multiple lines of code, was assessed using Parsons problems~\citep{parsons2006parson,ericson2022parsons} on questions about the same size, in terms of function size, as typical code writing exercises in the course. Finally, code explaining skill was assessed using Explain in Plain English (EiPE) exercises, presenting partially obfuscated code and requesting a high-level English description of what the code does~\citep{fowler2021should}.

Exams in this course were computer-based using \pl. All of the questions except for EiPE questions provided students with multiple attempts, with reduced credit for each incorrect answer, and used autograding to provide students with instant feedback while taking the exam. EiPE questions were manually scored by course staff after each exam. For each EiPE question, two course staff members individually graded the question as correct or incorrect. A third member provided their input in case of disagreements. 

\begin{table}[ht]
    \centering
    \small
    \begin{tabular}{|c|c|cccc|cccc|}
    \hline
    \multicolumn{2}{|c|}{Exam}& \multicolumn{4}{c|}{Write} & \multicolumn{4}{c|}{Trace} \\ \hline
     &  & Q1 & Mean & Median & Q3 & Q1 & Mean & Median & Q3\\
    \hline
    Exam1 & list & 0.5 & 0.73 & 0.92 & 1 & 1 & 0.99 & 1 & 1 \\
    & arithmetic & 1 & 0.96 & 1 & 1 & 1 & 0.95 & 1 & 1 \\
    & conditional & 0.62 & 0.81 & 0.96 & 1 & 1 & 0.94 & 1 & 1 \\
    & string & 0 & 0.66 & 0.86 & 1 & 1 & 0.9 & 1 & 1 \\
    \hline
    Exam2 & conditional & 0.66 & 0.76 & 0.86 & 0.98 & 1 & 0.94 & 1 & 1 \\
    & loop & 0.48 & 0.65 & 0.78 & 1 & 0.5 & 0.78 & 1 & 1 \\
    \hline
    Exam3 & pattern & 0.59 & 0.74 & 0.90 & 1 & 1 & 0.82 & 1 & 1 \\
    & advance string & 0 & 0.63 & 0.89 & 1 & 1 & 0.87 & 1 & 1 \\
    \hline
    Exam4 & loop &  &  &  &  & 1 & 0.87 & 1 & 1 \\
    & pattern & 0.83 & 0.82 & 0.95 & 1 & 1 & 0.89 & 1 & 1\\
    & dictionary & 0 & 0.54 & 0.78 & 1 & 1 & 0.96 & 1 & 1 \\
    & file & 0 & 0.54 & 0.71 & 0.89 & 1 & 0.81 & 1 & 1 \\
    & object & 0.2 & 0.62 & 0.82 & 1 &  &  &  & \\
    \hline
  \end{tabular}
  \newline
  \vspace*{0.5 cm}
  \newline
  \begin{tabular}{|c|c|cccc|cccc|}
    \hline
    \multicolumn{2}{|c|}{Exam}& \multicolumn{4}{c|}{Reverse Trace}  & \multicolumn{4}{c|}{Sequence} \\ \hline
     &  & Q1 & Mean & Median & Q3 & Q1 & Mean & Median & Q3 \\
    \hline
    Exam1 & list & 0.83 & 0.79 & 1 & 1 & 0 & 0.32 & 0 & 0.75 \\
    & arithmetics &  1 & 0.92 & 1  & 1  &  &  &  & \\
    & conditionals & 1 & 0.97 & 1 & 1 & 0.25 & 0.54 & 0.5 & 1 \\
    & string & 0.5 & 0.77 & 1  & 1 &  &  &  & \\
    \hline
    Exam2 & conditionals & 1 & 0.96 & 1  & 1 & 0.75 & 0.75 & 1 & 1 \\
    & loops & 0.25 & 0.58 & 0.5 & 1 & 0 & 0.22 & 0 & 0.25 \\
    \hline
    Exam3 & patterns & 0.50 & 0.73 & 1 & 1 & 0.81 & 0.83 & 1 & 1 \\
    & advanced strings & 1 & 0.80 & 1 & 1 & 0.83 & 0.86 & 1 & 1 \\
    \hline
    Exam4 & loops & 1 & 0.83 & 1 & 1 &  &  &  & \\
    & patterns & 1 & 0.82 & 1  & 1 &  &  &  & \\
    & dictionaries & 1 & 0.91 & 1 & 1 & 0.22 & 0.58 & 0.68 & 1 \\
    & files & 0.5 & 0.70 & 1 & 1 & 0.12 & 0.52 & 0.52 & 0.94 \\
    & objects & &  &  & & 0.12 & 0.56 & 0.67 & 1 \\
    \hline
  \end{tabular}
  \caption{Statistics drawn from the student score in each exams for each topics. Most of the distribution of the scores are left-skewed.}
  \label{tab:score_stats}
\end{table}

Table~\ref{tab:score_stats} shows the statistics of the student's grades (first quartile, mean, median, and third quartile) of four programming skills: write, trace, reverse trace, and sequence in each of the four exams. The first quartile (Q1), median (Q2), and third quartile (Q3) visualize the distribution of the dataset and provide the dataset information of the top 25\%, 50\%, and 75\%. Mean helps us to determine the average score in all those exams for each skill. 

Given that EiPE questions were manually graded, with only one attempt, these questions had discrete values only. There are empty boxes for some topics in Table~\ref{tab:score_stats} that indicate where certain topics were not assessed for specific skills.

\begin{landscape}
\begin{figure}[ht]
  \centering
  \includegraphics[width=1\linewidth]{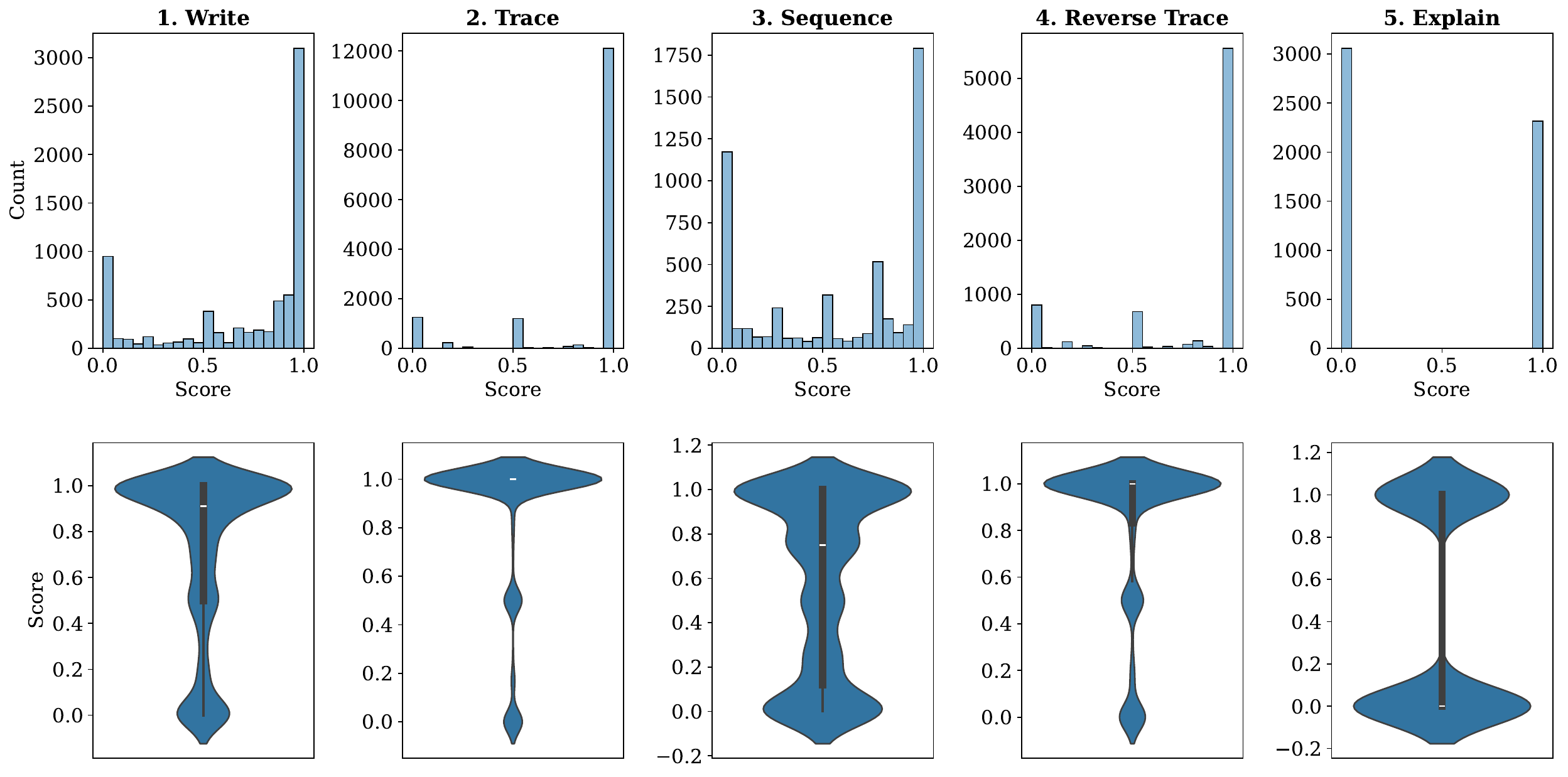}
   \caption{Student score distribution for all five skills. For explain, the scores are in discrete format. The rest four skills have continuous student score ranging from 0 to 1. The white dot in the box of the violin plot denotes the median of each of the five skills. The bar shows the quartile scores. }
  \label{fig:raw_scores}
\end{figure}
\end{landscape}

Figure~\ref{fig:raw_scores} shows the distribution of the student scores in all four exams. Most students performed excellently in four of the skills, with explaining code as the exception. The distribution of scores of four of the skills, write, trace, reverse trace and sequence, are left skewed with most students getting the answer correct on their first try.

The dataset was collected under the IRB number Anon.

\subsection{Data Analysis}
\subsubsection{Conviction as a prerequisite Measure}
As our measure of whether one skill is a prerequisite skill of another, we use the \textit{ Conviction } measure from association rule mining~\citep{kumbhare2014overview,brin1997dynamic}.  
For Skills A and B, Conviction is defined as
\begin{equation}
\textit{Conviction}(A \Rightarrow B)  = \frac{P(A )\cdot P(\neg B)}{P(A \text{ and } \neg B)}
\label{eq:conviction}
\end{equation}

Conviction can be thought of as intuitively similar to (though not as strict) as the logical implication $A \Rightarrow B$ or $A$ implies $B$. The need for a directional measure that goes beyond correlation was the motivator for the creation of the conviction measure~\citep{brin1997dynamic}, which is exactly what was needed to push forward research into the programming skill hierarchy~\citep{fowler2022reevaluating}. The idea is that if having skill~$A$ implies that a student also has skill $B$, then a student must have $B$ in order to have $A$, so skill~$B$ must have been a prerequisite for skill~$A$. Thus, if $\textit{Conviction}(A \Rightarrow B)$ is higher than $\textit{Conviction}(B \Rightarrow A)$, then skill $B$ is a prerequisite skill for skill~$A$.

Because conviction handles only binary variables, we convert our test data to binary variables by letting a student score be 1 if it is above a given threshold, or 0 if it is below or equal to the threshold (details on threshold selection are given in Section~\ref{sec:thresholding}).
So, in terms of our data set, the conviction is defined as

\begin{align}
\textit{Conviction}&(A \Rightarrow B) = \nonumber \\
&\frac{(\% \text{ above threshold Skill }A)\cdot (\% \text{ threshold or below Skill } B)}{(\% \text{ above threshold Skill } A \text{ and threshold or below Skill } B)}.
\label{eq:conviction-applied}
\end{align}

Conviction is a particularly useful choice as a measure of one skill being prerequisite for another because it does not only look at the correlation between two skills, but instead examines the likelihood of getting one of them correct given the other, controlling for the overall occurrence of both.
Examining the equation for Conviction helps us confirm that it measures what we want. If more students got skill~$A$ correct and $B$ wrong, we would think $B$ was not a prerequisite for $A$. In that case, the denominator decreases, making the overall conviction increase, which is what we would expect. 

Consider an illustrative example. In Table~\ref{tab:convict_example} are the number out of 100 imaginary students who scored above and below the threshold value for two arbitrary skills \textit{A} and \textit{B}. 

\begin{table}[b]
\centering
\begin{tabular}{l|llll}
   & A+ & A- &  &  \\ \cline{1-3}
B+ & 45 & 20 &  &  \\
B- & 5  & 30 &  &  
\end{tabular}
\caption{Conviction example table for 100 imaginary students. The + symbol indicates above the threshold and the - symbol indicates at or below the threshold. Out of 100 students, 50 got above the threshold on A, 50 got below the threshold on A, 65 got above the threshold on B, and 35 got below the threshold on B. .}
\label{tab:convict_example}
\end{table}

Given this, the \textit{Conviction} calculations for $\textit{Conviction}(A \Rightarrow B)$ and $\textit{Conviction}(B \Rightarrow A)$ are as follows:

\begin{align*}
\textit{Conviction}&(A \Rightarrow B) =
\frac{(0.5)\cdot (0.35)}{0.05} = 3.50 \\
\textit{Conviction}&(B \Rightarrow A) =
\frac{(0.65)\cdot (0.5)}{0.2} = 1.625
\end{align*}

Here, there is notably stronger evidence for \textit{B} to be a prerequisite of \textit{A}, which we would expect from Table~\ref{tab:convict_example}. There were 5 students that can perform on skill \textit{A} who were unable to perform on \textit{B}. While mastery of \textit{B} does not guarantee performance on \textit{A}, as shown by the 20 students who do not succeed, doing well on \textit{A} almost always means a student will succeed on the likely prerequisite, \textit{B}.

Prior work by ~\citet{pelchen2020evidence} used the Likelihood of Sufficiency (LS) and Likelihood of Necessity (LN), defined in earlier work by Duda and collaborators \citep{duda1976subjective,duda1981model}, as measures for prerequisite skills.
The LS is an extremely similar measure to Conviction---in fact, the relationship that holds between them is:
\begin{align}
\textit{LS} 
=& \frac{P(A | B)}{P(A | \neg B)} 
= \frac{P(A \text{ and } B)}{P(B)} \cdot \frac{P(\neg B)}{P(A \text{ and } \neg B)} \nonumber \\
=& \frac{\textit{Conviction}(A \Rightarrow B)}{Conviction(A \Rightarrow \neg B)}
\end{align}

Given the similarity between the two measures, we choose to use Conviction rather than LS because it is a simpler measure overall, and because it is more widely used in the literature. Re-running our analyses with LS rather than Conviction gives equivalent results.

The LN can also be written in terms of conviction:
\begin{align}
\textit{LN} 
=& \frac{P(\neg A | B)}{P(\neg A | \neg B)} 
= \frac{P(\neg A \text{ and } B)}{P( B)} \cdot \frac{P(\neg B)}{P(\neg A \text{ and } \neg B)} \nonumber \\
=& \frac{\textit{Conviction}(\neg A \Rightarrow B)}{Conviction(\neg A \Rightarrow \neg B)}
\end{align}

We choose not to use the LN measure for two reasons. The first being that when measuring if one skill is prerequisite to one another, it is important to use the same metric in both directions. If we use a different metric in different directions, then we are making some (potentially unwarranted) assumptions about the direction of the prerequisite relationship. The second reason is that the \textit{LN} uses students getting skill~$A$ incorrect as its primary measure. For open-ended questions like those in our data set, getting a skill incorrect is a very noisy signal, as students could have gotten questions incorrect because they did not know the material, or because they made a mistake. On the contrary, conditioning on getting a skill correct is more accurate---if a student got a skill correct then they most definitely have mastered that skill because guessing the correct answer on open-ended questions is near impossible.








\subsubsection{Selecting Thresholds for Conviction}
\label{sec:thresholding}

Conviction is only defined for binary values, but our dataset has continuous student exam scores except for Explain skills. So, to transform the data to binary, we need to pick some threshold. To make our analysis more robust, we re-run the entire analysis with multiple choices of threshold. We use mean, median, and third quartile (Q3) as the thresholds. For each topic in all five skills: write, trace, sequence, reverse trace, and explain, we calculated the threshold, and if the student score was higher than the calculated threshold value, we assigned it to 1, else 0. 

The following equation describes the thresholding strategy for the mean threshold for a given data point $x$ within the score distribution $X$:

\begin{equation}
f(x, X) = \begin{cases}
    1 \text{ if } x > \text{mean}(X) \\
    0 \text{ if } x \le \text{mean}(X)
\end{cases}
\end{equation}

There are some questions where the median score was either 0 or 1, requiring some special-casing when using the median threshold. Outside of special-casing, we assign 1 to scores greater than the median threshold. In the special-case where the median score \textit{is} 1, we assign 1 to scores greater than \textit{or equal} to the median, i.e., when the median score is 1. The following equation describes the thresholding strategy for the median threshold for a given data point $x$ within the score distribution $X$:

\begin{equation}
f(x, X) = \begin{cases}
    1 \text{ if } x = 1 \\
    1 \text{ if } x < 1 \text{ and } x > \text{median}(X) \\ 
    0 \text{ if } x < 1 \text{ and } x \le \text{median}(X)
\end{cases}
\end{equation}

Q3 is qualitatively similar to the median threshold, with the same special-casing. The following equation describes the thresholding strategy for the Q3 threshold for a given data point $x$ within the score distribution $X$:

\begin{equation}
f(x, X) = \begin{cases}
    1 \text{ if } x = 1 \\
    1 \text{ if } x < 1 \text{ and } x > \text{Q3}(X) \\ 
    0 \text{ if } x < 1 \text{ and } x \le \text{Q3}(X)
\end{cases}
\end{equation}

The statistical analysis of using Q3 as a threshold is similar to the median, so we present the results only for mean and median.


We had to drop two questions from the sequence vs. trace and explain vs. trace comparisons when using the median as a threshold because every student who scored above the median on the sequence and the explain question also scored above the median on the trace question, leading to a zero in the denominator.
This makes ${P(Sequence \text{ and } \neg Trace)}$ and ${P(Explain \text{ and } \neg Trace)}$ (the denominators of Equation~\ref{eq:conviction-applied}) 0, leading to an undefined.  These two instances are the only cases with an undefined conviction in our data set.
In this case, dropping the value has no impact on our results because we already have strong evidence that trace is a prerequisite to sequence and explain. Also, for one question, all students who scored above the median on sequence had scored above the median on the trace supports our decision. Similarly, with mean as a threshold, we had to drop the same two questions from our conviction distribution, and for the above reasoning, we dropped the values as it has no impact on the results.

\subsubsection{Conviction Distributions}


Because students in our data set completed similar programming tasks across many programming topics over the course of the semester, we were able to take many measurements of each of the programming skills that we studied. Because of this, rather than comparing students performance on a single measurement of a given skill as some prior work did~\citep{lopez2008relationships,pelchen2020evidence}, we are able to compare many measurements of a given skill to many measurements of another skill, enabling us to perform a statistical test in order to compare distributions rather than just individual skill pairs.

To test if there is a prerequisite relationship between two skills, Skill~$A$ and Skill~$B$, we follow this procedure: 
\begin{itemize}
\item Calculate $\textit{Conviction}(A \Rightarrow B)$ for all topics, giving us a distribution of values showing how strongly proficiency in Skill~$A$ suggests proficiency Skill~$B$
\item Calculate $\textit{Conviction}(B \Rightarrow A)$ for all topics, giving us a distribution of values showing how strongly proficiency in Skill~$B$ suggests proficiency in Skill~$A$
\item Perform a 2-sided Wilcoxon-rank sum test with the null hypothesis that neither of the skills are prerequisite for one another
\begin{itemize}
    \item if the $p-$value is below the significance threshold and the median of the  $\textit{Conviction}(A \Rightarrow B)$ distribution is greater than the median of the $\textit{Conviction}(B \Rightarrow A)$ distribution, then proficiency in Skill~$A$ suggests proficiency in  Skill~$B$ more strongly than Skill~$B$ suggests proficiency in Skill~$A$. This means that a student with Skill A must have already learned Skill~$B$, and so Skill~$B$ is a prerequisite for Skill~$A$
    \item if the $p-$value is below the significance threshold and the median of the  $\textit{Conviction}(B \Rightarrow A)$ distribution is greater than the median of the $\textit{Conviction}(A \Rightarrow B)$ distribution, then proficiency in Skill~$B$ suggests proficiency in Skill~$A$ more strongly than proficiency in Skill~$A$ suggests proficiency in Skill~$B$. This means that a student with Skill~$B$ must have already learned Skill~$A$, and so Skill~$A$ is a prerequisite for Skill~$B$
\end{itemize}    
\end{itemize}

We chose the Wilcoxon-rank sum test because most of the conviction distributions were non-normal. Note that our procedure does not make any assumptions about the ordering of skills, which differentiates our work from prior work~\citep{lopez2008relationships,pelchen2020evidence,yamamoto2011relationship}, which performed analyses with the assumption that code writing should be the ultimate or final skill to learn, and so did not check they hypothesis that code writing could itself be a prerequisite for another skill. 

We perform this procedure for all 10 pairs of the 5 skills we consider.
Because we are doing multiple comparisons, we readjust the threshold ($\alpha$) to reject the null hypothesis with the Bonferroni correction. Bonferroni correction prevents the rare events for multiple hypotheses with the equation $\alpha / m$. We changed the threshold $\alpha$ from 0.05 to 0.005; $m$ is 10 for our analysis with ten skills pair. Other methods such as the Benjamini-Hochberg procedure have also been used to correct for multiple comparisons~\citep{benjamini1995controlling}. In order to be conservative in claiming a relationship between skills that may not exist, we use the Bonferroni correction because it favors reducing false positives at the expense of having more false negatives. The Benjamini-Hochberg procedure would be inappropriate for this use case, as it prioritizes reducing false negatives and would introduce more false positives.

\section{Results}

To calculate the conviction, we paired the common topic for each skill pair and used the updated binary score using mean and median as a threshold. 
For each pair of skills, the distribution chart was created by treating each topic's conviction as a different distribution point.
We calculated the $W-$statistics and $p-$value using the score distribution for all ten pairs as shown in Table~\ref{tab:mean_result} and Table~\ref{tab:median_result}. The negative $W-$statistics indicate that Skill~$A$ is a prerequisite to Skill~$B$. Table~\ref{tab:mean_result} and Table~\ref{tab:median_result} are sorted in ascending order based on the $p-$value that indicates if one skill is a prerequisite of the other.

\begin{table}[ht]
    \centering
    \begin{adjustbox}{width=1\textwidth}
    \small
    \begin{tabular}{cccccccccc}
    \toprule
    \multicolumn{2}{c}{Skills} & \multicolumn{3}{c} {\textit{Conviction}($A \Rightarrow B$)} & \multicolumn{3}{c} {\textit{Conviction}($B \Rightarrow A$)} & \multicolumn{2}{c}{Comparison}\\
     Skill A   & Skill B   & Q1 & median & Q3 & Q1 & median & Q3 & $W-$statistic & $p-$value\\
    \midrule 
    trace & write & 1.04 & 1.06 & 1.08 & 1.17 & 1.5 & 1.7 & -3.51 & \signif{0.0004429472952323059} \\
    trace & explain & 1.02 & 1.02 & 1.04 & 1.22 & 1.34 & 1.49 & -3.36 & \signif{0.0007775304469403846} \\
    reversetrace & explain & 1.03 & 1.09 & 1.12 & 1.15 & 1.39 & 1.56 & -3.31 & \signif{0.0009285334224661572} \\
    write & explain & 1.16 & 1.19 & 1.25 & 1.43 & 1.58 & 1.84 & -3.22 & \signif{0.0012684282828717209} \\
    trace & seq & 1.03 & 1.03 & 1.1 & 1.26 & 1.3 & 1.43 & -2.87 & \signif{0.004040984683985582} \\
    trace & reversetrace & 1.06 & 1.08 & 1.15 & 1.13 & 1.22 & 1.33 & -2.6 & \signif{0.009374768459434895} \\
    reversetrace & write & 1.06 & 1.11 & 1.18 & 1.16 & 1.25 & 1.37 & -1.74 & \signif{0.08183746008958778} \\
    reversetrace & seq & 1.05 & 1.11 & 1.17 & 1.15 & 1.24 & 1.55 & -1.68 & \signif{0.09289194088370532} \\
    seq & explain & 1.09 & 1.18 & 1.31 & 1.24 & 1.4 & 1.43 & -1.26 & \signif{0.20757844233562417} \\
    write & seq & 1.17 & 1.24 & 1.44 & 1.22 & 1.37 & 1.63 & -0.66 & \signif{0.5078006482752733} \\
    \bottomrule
    \end{tabular}
    \end{adjustbox}
      \caption{
      Summary of results using Mean as a Threshold to calculate conviction. Because distributions were non-normal, we give the 
      Median and first and third quartile scores of the conviction distributions as summary statistics. Then we show Wilcoxon rank-sum test results of comparing the distributions. Tracing is a prerequisite to all four other skills, while writing is a prerequisite to explaining code. There is not a prerequisite relationship between sequencing and writing code.}
    \label{tab:mean_result}
\end{table}

\begin{figure*}[h]
  \centering
  \includegraphics[width=1\linewidth]{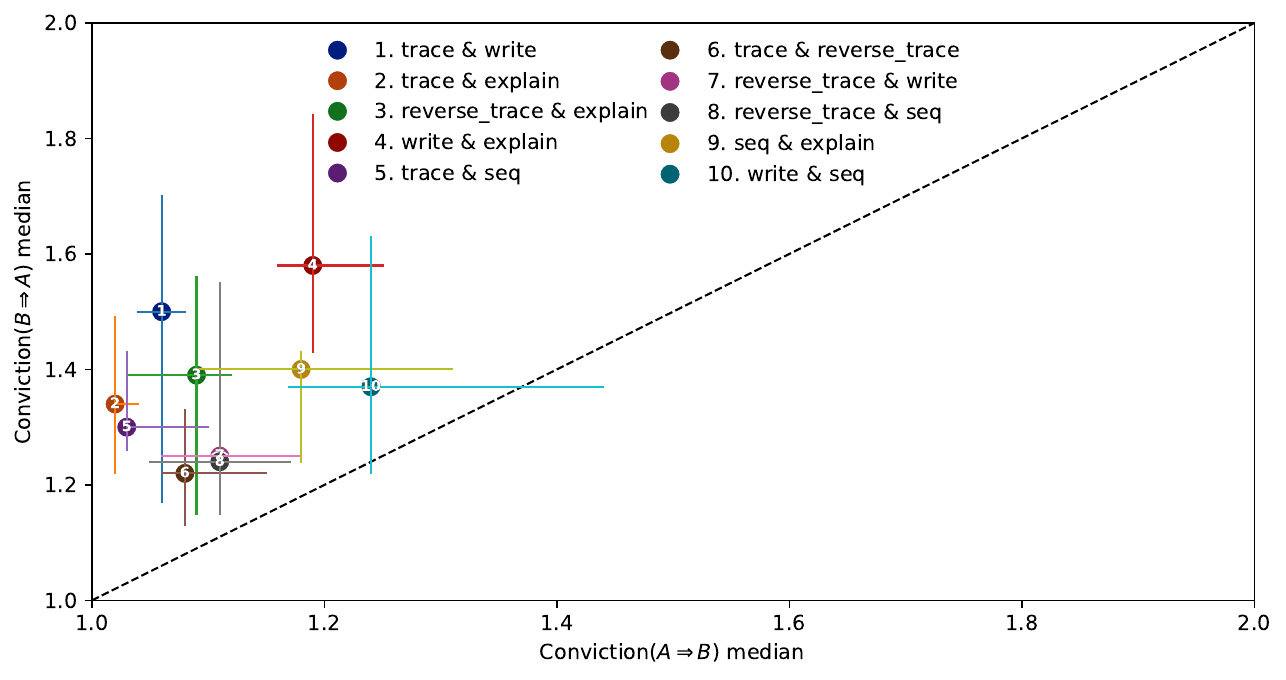}
  \caption{Visualization of conviction distributions for skills with conviction score calculated using mean threshold. Dots show medians, while the error bars shows the Q1 and Q3 score from the median. A point being close to the origin $(1,1)$ would represent no relationship between skills. A dot close to the y-axis, away from the origin implies a strong prerequisite relationship, since in that case $\textit{Conviction}(A \Rightarrow B)$ is near 1 while $\textit{Conviction}(B \Rightarrow A)$ is high. A dot midway between the dashed line and the y-axis represents a weaker prerequisite relationship. Dots close to the dashed line denote skills that are correlated but neither is a prerequisite of the other.}
  \label{fig:analysis_mean}
\end{figure*}

Our null hypothesis $H_{0}$ is that programming skill pairs are \textit{independent} of each other. If the p-value is less than 0.005 (after using Bonferroni correction), we reject the null hypothesis and conclude that a skill is a prerequisite to another. From both tables, Table~\ref{tab:mean_result} and Table~\ref{tab:median_result}, we found that trace is a prerequisite to the explain, sequence, and write skills while a reverse trace is a prerequisite to explain. Write is a prerequisite for the explain skill when using the mean threshold for data conversion, as shown in Table~\ref{tab:mean_result}. But we could not reject the null hypothesis for \texttt{write} $\leftrightarrow$ \texttt{explain} in the case where we use the median as a threshold for data conversion. Considering most of the data are left skewed, the median threshold value is greater than the mean for most of the skills as shown in Table~\ref{tab:score_stats}. This suggests that as students improve their overall programming ability, code writing ability and explaining ability tend to grow in tandem (see Table~\ref{tab:median_result}).

\begin{table}[ht]
    \centering
    \begin{adjustbox}{width=1\textwidth}
    \small
    \begin{tabular}{cccccccccc}
    \toprule
    \multicolumn{2}{c}{Skills} & \multicolumn{3}{c} {\textit{Conviction}($A \Rightarrow B$)} & \multicolumn{3}{c} {\textit{Conviction}($B \Rightarrow A$)} & \multicolumn{2}{c}{Comparison}\\
     Skill A   & Skill B   & Q1 & median & Q3 & Q1 & median & Q3 & $W-$statistic & $p-$value\\
    \midrule 
    trace & write & 1.02 & 1.05 & 1.07 & 1.32 & 1.49 & 2.05 & -3.84 & \signif{0.00012233301318702674} \\
    trace & explain & 1.02 & 1.02 & 1.04 & 1.22 & 1.34 & 1.49 & -3.36 & \signif{0.0007775304469403846} \\
    reversetrace & explain & 1.03 & 1.09 & 1.12 & 1.15 & 1.33 & 1.53 & -3.31 & \signif{0.0009285334224661572} \\
    reversetrace & write & 1.04 & 1.07 & 1.13 & 1.3 & 1.35 & 1.53 & -3.18 & \signif{0.001448775966454466} \\
    trace & seq & 1.03 & 1.03 & 1.08 & 1.28 & 1.35 & 1.46 & -3.13 & \signif{0.001745118699528905} \\
    trace & reversetrace & 1.06 & 1.08 & 1.15 & 1.08 & 1.2 & 1.33 & -2.31 & \signif{0.02092133533779403} \\
    reversetrace & seq & 1.05 & 1.09 & 1.13 & 1.11 & 1.29 & 1.48 & -2.1 & \signif{0.03569190011680441} \\
    seq & explain & 1.13 & 1.19 & 1.31 & 1.2 & 1.32 & 1.39 & -0.84 & \signif{0.40081416938293446} \\
    seq & write & 1.21 & 1.28 & 1.29 & 1.23 & 1.29 & 1.35 & -0.84 & \signif{0.4015419876158449} \\
    write & explain & 1.21 & 1.24 & 1.35 & 1.18 & 1.28 & 1.47 & -0.35 & \signif{0.7239320396139757} \\
    \bottomrule
    \end{tabular}
    \end{adjustbox}
    \caption{
      Summary of results using Median as a Threshold to calculate conviction. Because distributions were non-normal, we give the 
      Median and first and third quartile scores of the conviction distributions as summary statistics. Then we show Wilcoxon rank-sum test results of comparing the distributions. In contrast to mean threshold,  There is not a prerequisite relationship between explaining and writing code.}
    \label{tab:median_result}
\end{table}

\begin{figure*}[h]
  \centering
  \includegraphics[width=1\linewidth]{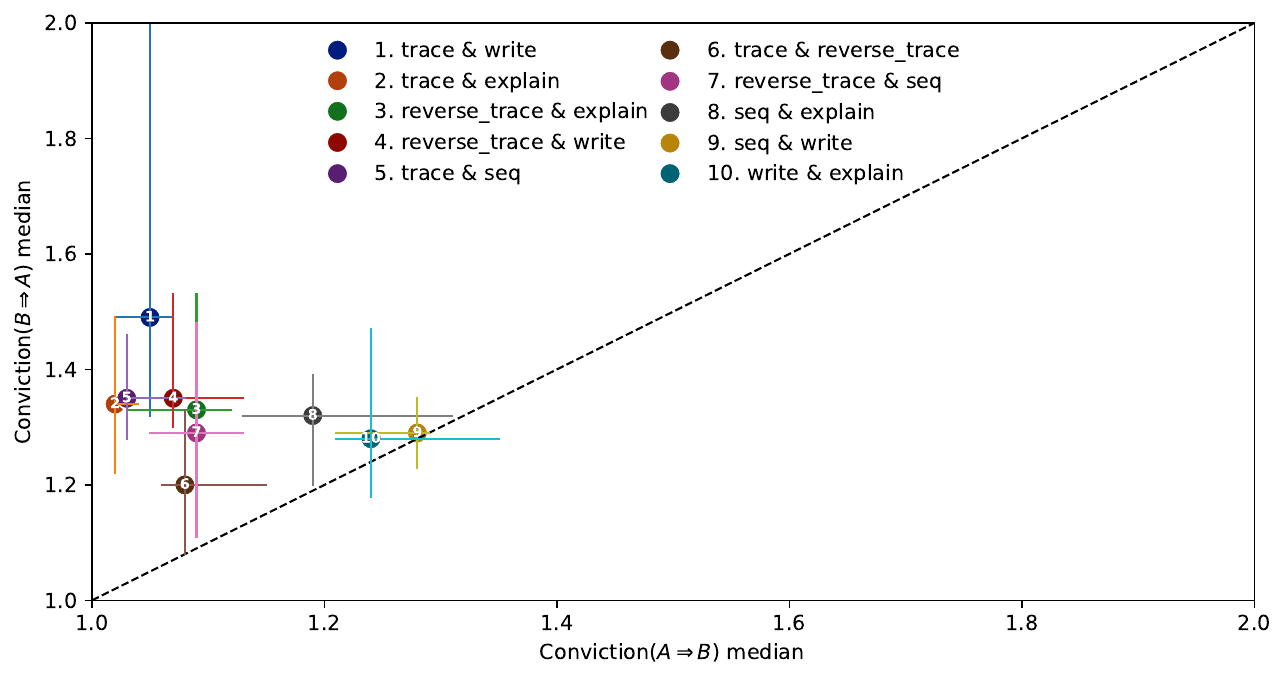}
  \caption{Visualization of conviction distributions for skills with conviction score calculated using median threshold. Dots show medians, while the error bars shows the Q1 and Q3 score from the median. A point being close to the origin $(1,1)$ would represent no relationship between skills. A dot close to the y-axis, away from the origin implies a strong prerequisite relationship, since in that case $\textit{Conviction}(A \Rightarrow B)$ is near 1 while $\textit{Conviction}(B \Rightarrow A)$ is high. A dot midway between the dashed line and the y-axis represents a weaker prerequisite relationship. Dots close to the dashed line denote skills that are correlated but neither is a prerequisite of the other.}
  \label{fig:analysis_median}
\end{figure*}

Plotting the median of $\textit{Conviction}(A \Rightarrow B)$ and $\textit{Conviction}(B \Rightarrow A)$ for all ten skill pairs, we got the points as shown in Figure~\ref{fig:analysis_mean} and Figure~\ref{fig:analysis_median}. Points away from the diagonal line show that the skill~$A$ is pre-requiste of skill~$B$. Two skills are not dependent with each other if the points are close to the diagonal line. From Table~\ref{tab:mean_result} and Table~\ref{tab:median_result}, we can dismiss the null hypothesis that two skills are independent with pairs with $p-$values less than $0.005$ (\num{5E-03}).

\section{Discussion}

\subsection{Strong evidence for tracing as a strict prerequisite skill, with explaining and writing close to terminal skills}

Our results give a strong argument for some prerequisite relationships. For trace-write, trace-explain, reverse trace-explain, and trace-sequence, there is a dependence among the skills using both mean and median as a threshold to convert data to binary. We do not reject the null hypothesis for reverse trace-sequence, sequence-explain, trace-reverse trace, and write and sequence ($p-$values and test statistics are shown in . There is a dependency of skill pair write-explain only using the mean as a threshold, while reverse trace-write has dependency when the median is used to convert data to binary. The graph structure shown in in Figure~\ref{fig:tree} shows if one skill is a prerequisite to another.

From our results, it follows that learning how to trace code first could help new programmers learn the other skills better. After learning how to trace, students can learn how to write code, reverse trace code, or sequence code segments. Students generally found explaining the code in plain English more challenging. From the tree Figure~\ref{fig:tree}, students who know how to trace, write, and reverse trace can efficiently learn to explain the code. If the instructor structures the courses in the assigned order of the hierarchy Figure~\ref{fig:tree}, it could help student learn the basics of programming skills better.

\subsection{Investigating co-requisiteness: some skills are even more tightly coupled than others}
While our analysis provides support from some clear prerequisite relationships, it also leaves open a question: given the interrelated nature of many of these skills, how do we identify \textit{co-requisite} skills? To that end, there is value at investigating the \textit{non-statistically significant} results. In particular, the relationships sequence-explain and write-sequence are interesting as their conviction score in either direction notably overlap. Their particularly high p-values for statistical significance combined with the fact that the conviction is high in both directions may instead be useful here to identify that the skills' performances are close enough to be \textit{co-requisite}.

In fact, this relationship makes some intuitive sense. Writing code naturally involves placing written lines of code in sequence. Likewise, when explaining code, the sequence of the lines of code will impact the code's behavior. Given that, it is not unreasonable to expect that increased proficiency with sequencing code may be beneficial to both writing and reading tasks. Relatedly, students who understand how to write code for a task or can explain code for a task may also be more likely to understand why the \textit{sequence} of code lines accomplishes that task.

\subsection{Relation to prior skill hierarchy work}
This work differs from prior work most significantly in that writing code was not found to be the capstone skill in the hierarchy. Of all existing hierarchies, only \textit{some} candidate hierarchies from \citet{fowler2022reevaluating} present the possibility of a write $\Rightarrow$ explain relationship and that work did not guarantee directionality. However, prior hierarchies \textit{assumed} write as the terminal skill and did not empirically validate that it \textit{was} the terminal skill. If we remove the apriori assumption that write is not a prerequisite skill for any others, then our findings are consistent with data presented in prior skill hierarchy work~\citep{lopez2008relationships,fowler2022reevaluating,yamamoto2011relationship}. 

It may be that this hierarchy is not the best way to consider these skills and that they are, instead, more naturally blended. \citet{xie2019theory} propose four skills: tracing, writing syntax, recognizing code templates, and using templates correctly. They propose that \textit{tracing} and \textit{template recognition} are reading skills, while \textit{writing syntax} and \textit{using templates} are writing skills. They abstract these out into reading and writing semantics, then reading and writing templates. Our findings are roughly consistent with this theory as well, although we note that \citet{xie2019theory} do not have an exact equivalent of explain. It may be that a final skill, \textit{communicating the use of templates}, may be a capstone fifth skill to add to the theory for programming instruction. Or, it could be the skill of code comprehension permeates all four and is revealed by exercises such as EiPE exercises.

    
    
    
    

\begin{figure*}[h]
  \centering
    \begin{tikzpicture}[
        > = stealth, 
        shorten > = 1pt, 
        auto,
        node distance = 3.5cm, 
        thick 
    ]
    
    \tikzstyle{state}=[
        draw = black,
        thick,
        fill = white,
        minimum width=25mm,
        minimum height=5mm
    ]
    
    \node[state] (trace) {Trace};
    \node[state] (sequence) [right of=trace, yshift=1cm] {Sequence};
    \node[state] (reverse_trace) [below of = sequence, yshift=1cm] {Reverse Trace};
    \node[state] (write) [right of=trace] {Write};
    \node[state] (explain) [right of = write] {Explain};
    
    \path[->] (reverse_trace) edge node {} (explain);
    \path[->, bend left=55] (trace) edge node {} (explain);
    \path[->, dashed] (write) edge node {} (explain);
    \path[->] (trace) edge node {} (write);
    \path[->] (trace) edge node {} (sequence);
    \path[->, dashed] (reverse_trace) edge node {} (write);
    
    \end{tikzpicture}
    \caption{Potential hierarchy Structure of the Programming Skills based on dependency. Solid line shows the prerequisiteness of a skill to a skill pointed with arrow for both threshold we used. Dotted line shows the prerequisiteness in only one of the threshold.}
    \label{fig:tree}
\end{figure*}

\section{Limitations and Future Work}
Our work has some threats to validity. First, we note that the grading of explain questions differs between hierarchies. Our explain questions were manually graded for either full credit or no credit along three dimensions: their technical accuracy, their unambiguity, and whether or not the answers reached the relational level of the SOLO taxonomy~\citep{lister2006not}. \citet{lopez2008relationships} allowed for partial credit in their grading, which could have created differences in how the explain skill was measured. However, given similar average scores (47\% in our work and 40\% in \citet{lopez2008relationships}), we find it unlikely that grading differences explain our different conclusions. As stated earlier, it is more likely that our difference in conclusions come from different starting assumptions---specifically, that we did not make the assumption that writing code could not be  a prerequisite skill for other skills.

A second threat to validity comes from the grading methods our exams used. Our questions for all but explain were autograded and provided instant feedback to students during exam conditions. This could mean that students show higher performance on some question types than they would in a manually graded setting, where many of the hierarchies were established. Further, this could mean that explain was more likely to be incorrect due to students not getting a chance to correct their mistakes. However, as we threshold whether or not students got a question correct or incorrect at the mean for each question, we mitigate the impact of this grading method on our modeling.

Finally, there is a clear limitation in what our work is showing. Our work shows a more supported hierarchy than prior work, but this hierarchy is \textit{not} a teaching prescription. Just because we can say ``if someone knows how to explain code, they probably already know how to write code,`` does \textit{not} mean that code writing must be taught before code reading. Future studies, as called for by \citet{fowler2022reevaluating}, should be carried out to explicitly validate teaching order. For example, while our analysis shows that sequence is not likely a prerequisite skill of write, there is work showing that sequencing code can help people learn to write code more quickly~\citep{ericson2022parsons}.
Nevertheless, our work may help inform which teaching orders to consider prior to such studies. For example, while a common assumption that writing is the capstone skill for teaching may still be valuable to assess, an order that either ends in or relies too heavily upon sequencing can likely be discarded. In particular, we think that more studies examining the relationship between the skills of writing code and explaining code would be particularly beneficial.

\section{Conclusion}
We studied the dependence of pairs of skills between five programming skills: writing, tracing, reverse tracing, sequencing, and explaining programming code. We converted the student score to binary using mean and median as thresholds and calculated the conviction score from association rule mining on the transformed datasets. Our findings show that for both thresholds, tracing is a prerequisite skill before learning how to write, explain, and sequence code. As the student improve their skill, writing code appears to become more co-requisite with explaining. Our findings show that  \texttt{seq} $\leftrightarrow$ \texttt{explain} and \texttt{write} $\leftrightarrow$ \texttt{seq} are co-dependent on each other. 
Using our findings, we created a skill hierarchy chart showing the prerequisites among skill pairs with stronger support than previous hierarchies. 

\appendix

\bibliographystyle{apacite}
\bibliography{references}

\end{document}